\documentclass[conference]{IEEEtran}
\usepackage{amsmath,amssymb,amsfonts,mathrsfs,graphicx,hhline,subcaption}
\usepackage[ruled,vlined,linesnumbered]{algorithm2e}
\usepackage{geometry}
\usepackage{booktabs} 
\usepackage{CJK}
\usepackage{enumitem,color}

\begin{document}

\title{\emph{RoNet}: Toward Robust Neural Assisted Mobile Network Configuration\vspace{-0.05in}}

\author{
\IEEEauthorblockN{Yuru Zhang, Yongjie Xue, Qiang Liu \\ University of Nebraska-Lincoln \\ qiang.liu@unl.edu\vspace{-0.4in}}
\and
\IEEEauthorblockN{Nakjung Choi \\ Nokia Bell Labs \\ nakjung.choi@nokia-bell-labs.com\vspace{-0.4in}}
\and
\IEEEauthorblockN{Tao Han \\ New Jersey Institute of Technology \\ tao.han@njit.edu\vspace{-0.4in}}
\vspace{-0.3in}}

\maketitle

\begin{abstract}
Automating configuration is the key path to achieving zero-touch network management in ever-complicating mobile networks.
Deep learning techniques show great potential to automatically learn and tackle high-dimensional networking problems.
The vulnerability of deep learning to deviated input space, however, raises increasing deployment concerns under unpredictable variabilities and simulation-to-reality discrepancy in real-world networks.
In this paper, we propose a novel \emph{RoNet} framework to improve the robustness of neural-assisted configuration policies.
We formulate the network configuration problem to maximize performance efficiency when serving diverse user applications.
We design three integrated stages with novel normal training, learn-to-attack, and robust defense method for balancing the robustness and performance of policies.
We evaluate \emph{RoNet} via the NS-3 simulator extensively and the simulation results show that \emph{RoNet} outperforms existing solutions in terms of robustness, adaptability, and scalability.
\end{abstract}


\vspace{-0.05in}
\section{Introduction}
\label{sec:introduction}
Emerging applications and services, e.g., augmented reality and autonomous driving, generate diversified and surging networking demands, e.g., latency and throughput, and reliability.
To meet the ever increasing traffic demands, next-generation mobile networks evolve toward open, ultra-dense and disaggregated~\cite{polese2022colo}, e.g., radio unit (RU), centralized unit (CU) and distributed unit (DU).
The complex mobile network requires intelligent and automated configuration policies to efficiently serve different users and groups.

Network configuration is the key mechanism to dynamically configure diverse network parameters~\cite{shi2021adapting}, e.g., bandwidth, spectrum and transmission mode, on various physical infrastructures such as base stations and core servers.
Different from fine-grained resource management, e.g., power and resource allocation, network configurations are enforced with larger time intervals, e.g., 30 minutes~\cite{marquez2018should} and hours, as they may involve sophisticated infrastructural operations.
The large configuration intervals weaken the temporal inter-dependencies between consecutive configuration actions, in other words, network configuration generally does not hold Markov property and can be recognized as multiple independent one-shot configuration problems~\cite{liu2019virtualedge, salvat2018overbooking}.

Existing approaches~\cite{ran2018deepdecision, d2020sl, d2022orchestran, salvat2018overbooking} rely on approximated mathematical models to represent the mobile network and use optimization based methods (e.g., linear and convex optimization) to optimize different performance metrics, e.g., throughput and energy consumption.
As configuration parameters expand to be high dimensional, e.g., hundreds if not more, these approaches fail to achieve accurate representation of mobile networks and fail in real-world deployment~\cite{liu2021onslicing, bega2019deepcog}.
Recent, deep learning techniques have been increasingly explored and adopted for high-dim network configuration in mobile networks.
For example, a deep neural network (DNN) can be instantiated and trained to approximate the correlations among observed network state, configuration parameters, and resulted network performances~\cite{shi2021adapting}. 
With the continual learning of ML models over continuously collected transitions, the ML-based approaches shed the light to achieve zero-touch management for next-generation mobile networks. 


In real-world mobile networks, the observation of high-dim network state can be deviated by unpredictable variabilities, e.g., inaccurate state collection, traffic fluctuations, link failure, and simulation-to-reality discrepancy (the difference between offline simulator and real-world networks~\cite{liu2021onslicing, shi2021adapting}).
The deviated observations may not be seen in the training dataset of DNN models previously, which usually results in the degradation of prediction accuracy~\cite{madry2017towards}, uncertain configuration actions, and compromised network performances.
Besides, potential adverse attackers may attack and manipulate the state observation~\cite{madry2017towards}, where the changes on state are not revealed to configuration policies.
As a result, configuration policies determine configuration actions based on pre-attack state, which may substantially degrades and even collapses network performances, e.g., long user latency and flow congestion.
Hence, it is imperative to investigate effective approaches to improve the robustness of ML-based configuration policies under diversified network uncertainties.

In this paper, we propose \emph{RoNet}, a new robust network configuration framework in mobile networks with three integrated stages.
We formulate the network configuration problem to maximize the performance efficiency (PE) under varying network states. 
First, we design a new neural-assisted network configuration policy to effectively address the problem.
Second, we design a novel learn-to-attack method to automatically attack state observation without the need for prior knowledge about the policy.
Third, we design a novel robust defense method to recover network performances of the policy.

To the best of our knowledge, we are the first work for assuring the robustness of neural-assisted configuration policies in mobile networks. 
The specific contributions are summarized as follows:
\begin{itemize}
    \item We formulate the network configuration problems to maximize the performance efficiency in mobile networks.
    \item We design a new \emph{RoNet} framework to improve the robustness of neural-assisted configuration policies.
    \item We design three integrated stages with new network configuration, learn-to-attack and robust defense methods. 
    \item We evaluate \emph{RoNet} via NS-3 simulator extensively and results justify the superiority of \emph{RoNet} in terms of robustness, adaptability and scalability. 
\end{itemize}

\vspace{-0.05in}
\section{System Model}
We consider a mobile network with multiple base stations in radio access networks (RAN), network switches in transport networks (TN), the core network (CN), and servers in edge networks (EN).
The mobile network is considered to be time slotted~\cite{marquez2018should}, where configuration actions can only be made in discrete time slots, e.g., every hour.
Without loss of generality, we consider that network configurations may be applied to individual mobile users, user groups (e.g., network slices~\cite{foukas2017orion}), or particular infrastructures (e.g., specific gNBs). 
For the sake of simplicity, we refer to user groups in the following descriptions.
At each configuration time slot, the operator can observe the network state by gathering a collection of statistical metrics, e.g., user traffic and transport network delay.
As the configuration action is enforced, the operator can obtain the performance of users, e.g., statistics of users, before the next configuration interval.


\textbf{State Space.}
The network state $s_t$ represents the current status of mobile networks in multiple aspects, and has the impact on the performance of mobile users.
We define the state space $s_t$ in the Table~\ref{tb:simulation_parameter_space}.
The parameters \emph{ul\_avg\_size} and \emph{dl\_avg\_size} are the average uplink and downlink data size of user applications. 
The parameters \emph{mcs\_max\_ul} and \emph{mcs\_max\_dl} are the maximum uplink and downlink modulation and coding scheme (MCS) of users. 
The parameter \emph{avg\_distance} represents the average geographic distance between users and base stations.
Here, the design of state space can be easily extended to support more parameters if applicable, e.g., scheduler algorithm and transmission power.

\textbf{Action Space.}
The configuration action $a_t$ allows a variety of network configuration in multiple technical domains to be enforced in mobile networks. 
We define the action space in the Table~\ref{tb:configuration_space}. 
The parameter \emph{bandwidth\_ul} and \emph{bandwidth\_dl} are the uplink and downlink wireless bandwidth of users, respectively.
The parameter \emph{cpu\_ratio} is the CPU share ratio of edge servers for serving user applications.
Here, the design of action space can incorporate more configuration parameters if available, e.g., the bandwidth allocation in transport networks.

\begin{table}[!t]
\centering
    \begin{tabular}[b]{c|c}\hline
       \textbf{States}                       &  \textbf{Meaning} \\ \hline
       \textbf{ \emph{ul\_avg\_size}}            &  average uplink data size of user applications  \\ 
       \textbf{ \emph{dl\_avg\_size}}            &  average downlink data size of user applications \\ 
       \textbf{ \emph{mcs\_max\_ul}}            &  maximum uplink modulation and coding scheme  \\ 
       \textbf{ \emph{mcs\_max\_dl}}            &  maximum downlink modulation and coding scheme  \\ 
       \textbf{ \emph{avg\_distance}}                &  average distance between users and base station  \\ \hline
    \end{tabular}
    \captionof{table}{The state space}
\label{tb:simulation_parameter_space}
\end{table}

\begin{table}[!t]
\centering
    \begin{tabular}[b]{c|c|c}\hline
       \textbf{Actions}                       &  \textbf{Meaning} & \textbf{Range} \\ \hline
       \textbf{ \emph{bandwidth\_ul}}        &  maximum uplink physical resource blocks  & [0, 50]\\ 
       \textbf{ \emph{bandwidth\_dl}}           & maximum downlink physical resource blocks & [0, 50] \\ 
       \textbf{ \emph{cpu\_ratio}}       & CPU share ratio of edge server  & [0, 1]\\ \hline    
    \end{tabular}
    \captionof{table}{ The configuration action space}
\label{tb:configuration_space}
\end{table}

\textbf{Performance Efficiency.}
Given the network state and configuration action, the performance of users can be retrieved at the end of individual configuration intervals, e.g., the collection of end-to-end latency.
Due to the complicated and underlying correlations in mobile networks~\cite{shi2021adapting, ayala2021bayesian}, the performance function is commonly considered to be unknown.
We denote the performance function as $f(s_t, a_t)$, which is related to the network state $s_t$ and configuration action $a_t$.
Due to the heterogeneity of user applications, their performance metrics can be extremely diversified, e.g., delay, throughput, and reliability.
Thus, we define a unified metric, i.e., PE, to evaluate how the mobile network satisfies the user needs
\begin{equation}
\label{eq:PE}
Q(s_t, a_t) = Prob(f(s_t, a_t)<H) / |a_t|,
\end{equation}
where $Prob(f(s_t, a_t)<H)$ represents the percentile performance and $H$ is the threshold~\cite{salvat2018overbooking}.
For example, if the user application is latency-oriented, e.g., augmented reality, the percentile performance means how many percent latencies are lower than the given threshold.
The $|a_t|$ represents resource usage of users in multiple technical domains, which is determined by the configuration action $a_t$.
To cost-efficiently serve diverse user applications, the network operator aims to strike the balance between percentile performance and resource usage.

\textbf{Problem.} 
Given a time period $T$, the objective is to derive the optimal policy ($\pi$) that maximizes the cumulative PE. Thus, we formulate the problem $\mathbb{P}_0$ as 
\begin{align}
     \mathbb{P}_0: \max \limits_{\pi} & \;\;\;\;\;  \underset{\pi}{\mathbb{E}}\left[ \sum\nolimits_{t=0}^{T} Q(s_t, a_t) \right] \label{eq:org_objective} \\ 
     s.t. & \;\;\;\;\; 0 \le a_t \le M, \label{eq:const_action} 
\end{align}
where $M$ are the maximum configuration actions, e.g., total downlink wireless bandwidth in Table~\ref{tb:configuration_space}.
Note that network states $s_t$ may change at different configuration intervals during the time period $T$.
The main challenge of resolving the problem lies in the complex and unknown performance function.
On the one hand, existing model-based convex optimization methods~\cite{boyd2004convex}, e.g., gradient descent, cannot be directly applied.
On the other hand, the high-dimensional state and action space prohibit existing searching methods, e.g., exhaustive and grid searching.

\vspace{-0.05in}
\section{Neural-Assisted Network Configuration}
\label{subsec:normal_training}
In the normal training stage, we propose a new \underline{n}eural \underline{a}ssisted \underline{n}etwork c\underline{o}nfiguration (NANO) method to effectively resolve the problem $\mathbb{P}_0$ in two steps.
First, we create a deep neural network (DNN) and train it to predict performance of users under both the state and action space.
Second, we develop a searching scheme to randomly sample actions from the action space and select the optimal configuration action that has the maximum predicted PE.

\textbf{Neural-Assisted Prediction.}
\label{subsubsec:neural_prediction}
Deep neural networks have been demonstrated very promising accuracy on complex function regression and prediction, e.g., computer vision and natural language processing.
We resort to DNN-based approaches to predict the performance of users, because conventional approaches, e.g., decision tree, fail to tackle high-dim state and action space with high prediction accuracy.
In particular, we create a DNN (denoted as $\pi_\theta$) with weights $\theta$, whose output is 1-dim to predict the PE of users. 
Its input includes both the state space $s_t$ and action space $a_t$ to be aware of current network state and configuration action.
To construct the training dataset, we collect the observed network states and sample the action space with grid searching from either network simulators or real-world networks.
We adopt stochastic gradient descent methods to train the DNN by adopting the mean squared error (MSE) loss function.

\textbf{Randomized Action Searching.}
\label{subsubsec:random_searching}
With the trained DNN for PE prediction, we develop a randomized action searching scheme to determine the optimal configuration action under different network states.
First, we randomly sample thousands of actions from the action space (3-dim in Table~\ref{tb:configuration_space}), where more samples may be needed when the dimension of action space increases.
Second, we concatenate the current network state and these sampled actions, feed them into the DNN model, and obtain the predicted PEs.
Third, we select the optimal configuration action that has the maximum predicted PE as follows
\begin{equation}
    a^*_t = \underset{a_t}{\arg\max} \;\; \pi_\theta(s_t, a_t).
\end{equation}



\section{The RoNet Framework}
\begin{figure}[!t]
\centering
\includegraphics[width=2.5in]{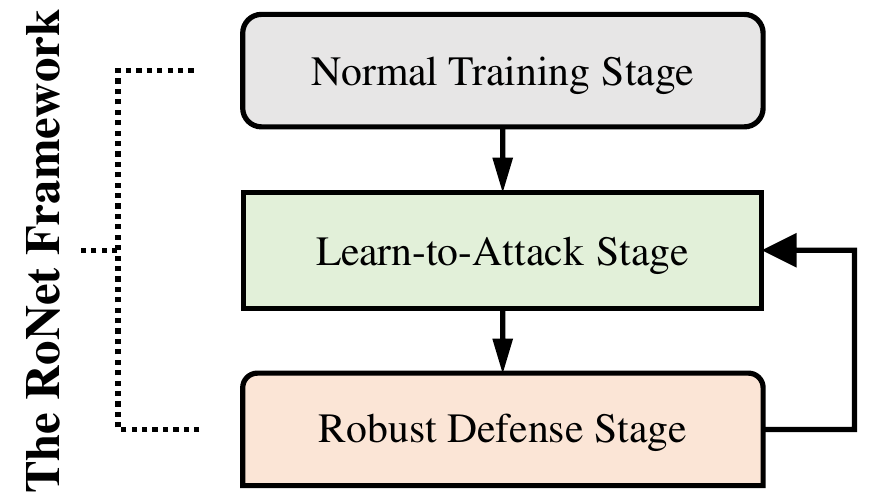}
\caption{\emph{RoNet} Overview}
\label{fig:architecture}
\end{figure}

In this section, we introduce the \emph{RoNet} framework in Fig.~\ref{fig:architecture} to improve the robustness of configuration policies.
First, we design a new learn-to-attack method to automatically learn and attack the neural-assisted configuration policy, without the need for prior knowledge, e.g., gradients.
Second, we design a new robust defense method to defend and recover the performance of the policy while it is under attack.
The above procedures may be repeated to adaptively strike the balance between the robustness and performance of the policy.

\subsection{Learn-to-Attack Stage}
\label{subsec:learn_to_attack}
Existing adversarial attack approaches commonly assume that the information of the model is known by the attacker, which allows adversarial gradient-based attacks.
However, the configuration policy is composed of not only the DNN model but also the action searching.
Existing approaches, that attack the accuracy of the DNN model, may not effectively degrade the overall performance of the configuration policy.
Thus, we design the attacks to the configuration policy without the assumption of known DNN model.

\textbf{Attacker Problem.}
The objective of the attacker (denoted as $\pi_v$) is to minimize the performance of the policy under the given scale of attacks.
Thus, we formulate the attacker problem as 
\begin{align}
     \mathbb{P}_1: \min \limits_{\pi_v} & \;\;\;\;\;  \underset{\pi_v}{\mathbb{E}}\left[\sum\nolimits_{t=0}^{T} Q(s_t + \pi_v(s_t), a_t) \right] \label{eq:atk_objective} \\ 
     s.t. & \;\;\;\;\; |\pi_v(s_t)|_\infty \le \epsilon, \label{eq:atk_const} 
\end{align}
where the $\pi_v(s_t)$ generates the attack by observing the current state $s_t$.
The attack space is defined by the maximum scale ($\epsilon$) with the regulation of the $l_\infty$-norm.

\textbf{Bayesian Learning.}
\label{subsubsec:policy_robustness}
Bayesian learning is a state-of-the-art approach to automatically learn and tackle complex unknown problems~\cite{ayala2021bayesian}.
It relies on a surrogate model to approximate the observable performance and an acquisition function to determine the next attack on the state ($\hat{s}_t$) to query.
In each iteration, the surrogate model will be trained with all collected transitions of attacks and degraded PE performance, and the acquisition function will be recalculated according to the updated surrogate model.
The maximization of the acquisition function will generate the next attack, which will be queried, e.g., simulator or real-world testbeds, to obtain the corresponding PE performance.

\textbf{Gaussian Process.}
Gaussian Process~\cite{rasmussen2003gaussian} has been extensively adopted as the surrogate model and shown great successes in a variety of practical problems.
For the sake of simplicity, we denote the observed dataset as $\mathcal{D}^{1:t}=\{\mathcal{V}^{1:t}, y^{1:t}\}$, where $\mathcal{V}^{1:t}$ and $y^{1:t}$ are the set of attacks and the corresponding negative PEs until the iteration $t$, respectively. 
The posterior distribution is expressed as $P(y|\mathcal{D}^{1:t}) \propto P(\mathcal{D}^{1:t}|y)P(y)$.
By using Gaussian Process $\mathcal{GP}$~\cite{rasmussen2003gaussian} as the prior, $y$ can be represented as $y \sim \mathcal{GP}(\mu, k)$.
$\mu(v)$ is the mean function and $k(v,v')=\exp(-\frac{l}{2}||v-v'||^2)$ is covariance function under the given attack $v$.
Here, we use the most common kernel function, i.e., radial basis function (RBF) with the lengthscale $l$ is set as 1.0.
Given an arbitrary attack $\tilde{v}$, the posterior distribution can be derived as
\begin{equation}\label{eq:predict_distribution}
	P(y(\tilde{v})|\mathcal{D}^{1:t},\tilde{v}) \sim \mathcal{N}(\mu(\tilde{v}),\sigma^2(\tilde{v})),
\end{equation}
where $\mu(\tilde{v}) = \mathbf{k}^T[\mathbf{K}+\delta^2 \mathbf{I}]^{-1}y^{1:t}$, and $
	\sigma^2_{i}(\tilde{v}) = k(\tilde{v},\tilde{v})-\mathbf{k}^T[\mathbf{K}+\delta_{noise}^2 \mathbf{I}]^{-1}\mathbf{k},$
where $\mathbf{k}=[k(\tilde{v}, v^{1}), k(\tilde{v}, v^{2}), \cdots, k(\tilde{v}, v^{t})]$ and $\mathbf{K}= [k(v^{i}, v^{j})]_{t\times t}, \forall i, j =1,2,...,t$. The $\mathbf{I}$ denotes the identity matrix with the same dimensions as $\mathbf{K}$.

\textbf{Acquisition Function.}
There are various candidate acquisition functions, e.g., EI, PI and UCB, can be used to select the next attack, according to the updated surrogate model.
Without loss of generality, we adopt the Gaussian process upper confidence bound (GP-UCB), which has been extensively evaluated to be robust in tackling diversified scenarios~\cite{srinivas2009gaussian}.
Given the generated mean $\mu$ and variance $\sigma^2$ function, GP-UCB selects the next action as follows
\begin{equation}
    v_{next} = \underset{v_t}{\arg\max} \;\; {\mu(v_t) + \sqrt{\beta_t} \cdot \sigma(v_t)},
\end{equation}
where $\beta_t$ is a non-negative hyperparameter to balance the exploration and exploitation.
Intuitively, the higher $\beta_t$ encourages more exploration, while the lower $\beta_t$ focuses on exploitation.
With the dedicated selection of $\beta$ in individual iterations, the application of GP-UCB acquisition function has a solid theoretical guarantee on converging to global optima~\cite{srinivas2009gaussian}.
In particular, under the selective kernel functions~\cite{srinivas2009gaussian}, the sub-linear regret is achieved with the probability of $1-\delta$, if 
\begin{align}
\label{eq:gp_ucb}
    \beta_t = 2{\log\left(\frac{t^2  \pi^2 }{3 \delta} \right)} + 2d{\log\left(t^2 d  \eta_2  r  \sqrt{\log(\frac{4d\eta_1}{\delta})} \right)},
\end{align}
where $\delta$ is the hyperparameter between 0 and 1, and $\eta_1>0, \eta_2>0, r >0$ are constants, and $d$ is the attack dimension.


\textbf{Remark.}
In this stage, the configuration policy is attacked by the learn-to-attack method by perturbing its observable state space.
As the attack completes, we can obtain the collection of network states, attack on states, configuration actions, and the attacked PE performance.

\vspace{-0.05in}
\subsection{Robust Defense Stage}
\label{subsec:robust_training}
To counter the attack and recover the performance, we design a new robust defense method that is composed of model retraining and probabilistic action selection.
The model retraining aims to recover the accuracy of the DNN model $\pi_\theta$ in terms of predicting the attacked PE performance.
The probabilistic action selection further improves the robustness by randomly choosing selective configuration actions.

\textbf{Model Re-Training.}
As the configuration policy is unaware of the existence of the adverse attacker, its DNN model observes the pre-attack states (instead of attacked states) and generates predicted PE performance.
We observe that this deviation between pre-attack and attacked states partially constitutes the performance degradation of the policy.
We retrain the DNN model based on the attacked dataset, i.e., the collection of pre-attack states and their corresponding attacked PE performance.
In particular, we design the loss function of $\pi_\theta$ retraining as 
\begin{equation}
    Loss_{\pi_\theta} = |\pi_\theta(s_t, a_t) - U(s_t + \pi_v(s_t), a_t)|_2,
\end{equation}
where $U(s_t + \pi_v(s_t), a_t)$ is the attacked PE performance under the network state $s_t$ and the attacker $\pi_v$.
The attack $\pi_v(s_t)$ is not revealed to the DNN model $\pi_\theta$.
As the model retraining completes, the DNN model $\pi_\theta$ will recover its prediction accuracy in terms of the attacked PE performance.

\textbf{Probabilistic Selection.}
In the normal training stage, the policy selects the configuration action by maximizing the predicted PE, which commonly results in non-robust performance in practice.
On one hand, the DNN model $\pi_\theta$ may not be very accurate on PE prediction under any states and actions.
On the other hand, the deterministic action selection under the given DNN model can be easily attacked by targeting the particular maximum.
Hence, we design the probabilistic selection scheme as follows.
First, we sample the action space extensively with thousands of samples.
Second, we concatenate the current state and these sampled actions, and feed them into the retrained DNN model.
Third, we rank all the predictions of PE performance and truncate the first $\kappa$th percentile.
Fourth, the configuration action is randomly selected from the truncated set.
In this way, the probabilistic action selection not only mitigates the effect of inaccurate PE prediction but also counters the potential attacks implicitly.

\textbf{Remark.}
In this stage, the performance of the configuration policy is recovered by using model retraining and probabilistic selection.
The learn-to-attack and robust defense stages may be repeated to strike the balance between the robustness and the performance of the policy.

\begin{figure*}[!t] 
\captionsetup{justification=centering}
  \begin{minipage}[t]{0.325\textwidth}
    \centering
    \includegraphics[width=2.3in]{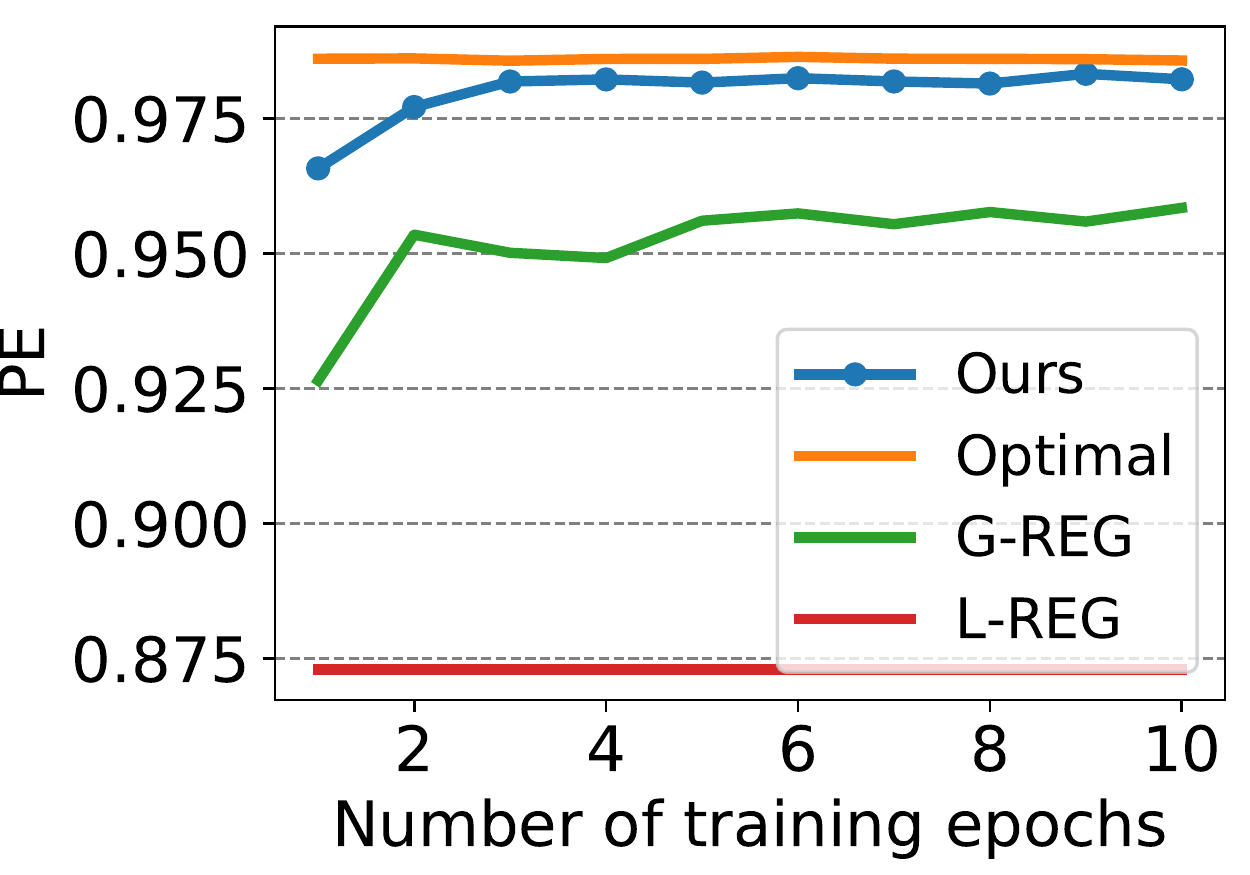}
    \captionof{figure}{Performance under normal training}
    \label{fig:result_normal_range}
  \end{minipage}
  \hfill
  \begin{minipage}[t]{0.325\textwidth}
    \centering
    \includegraphics[width=2.3in]{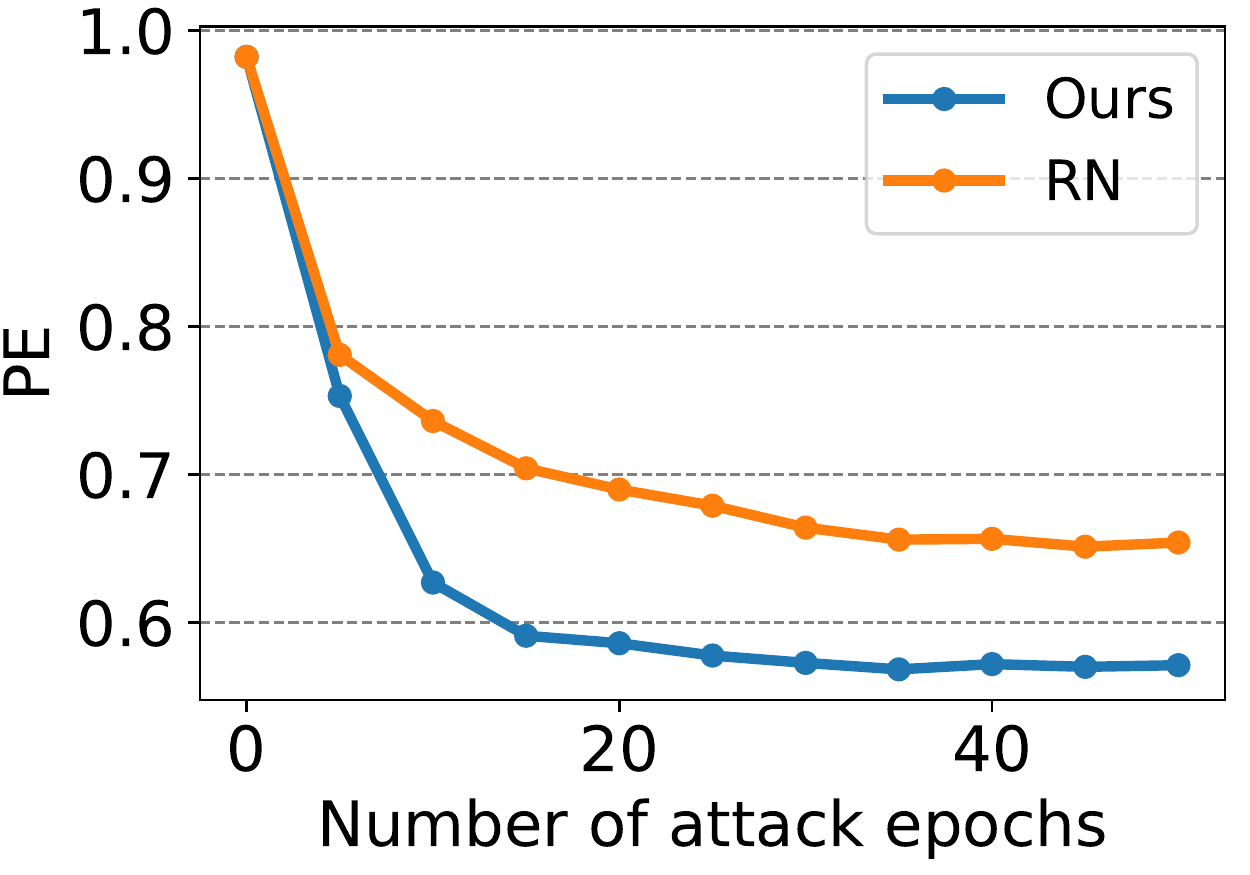}
    \captionof{figure}{Performance under attacks}
    \label{fig:result_attack_range}
  \end{minipage}
  \hfill
  \begin{minipage}[t]{0.325\textwidth}
    \centering
    \includegraphics[width=2.3in]{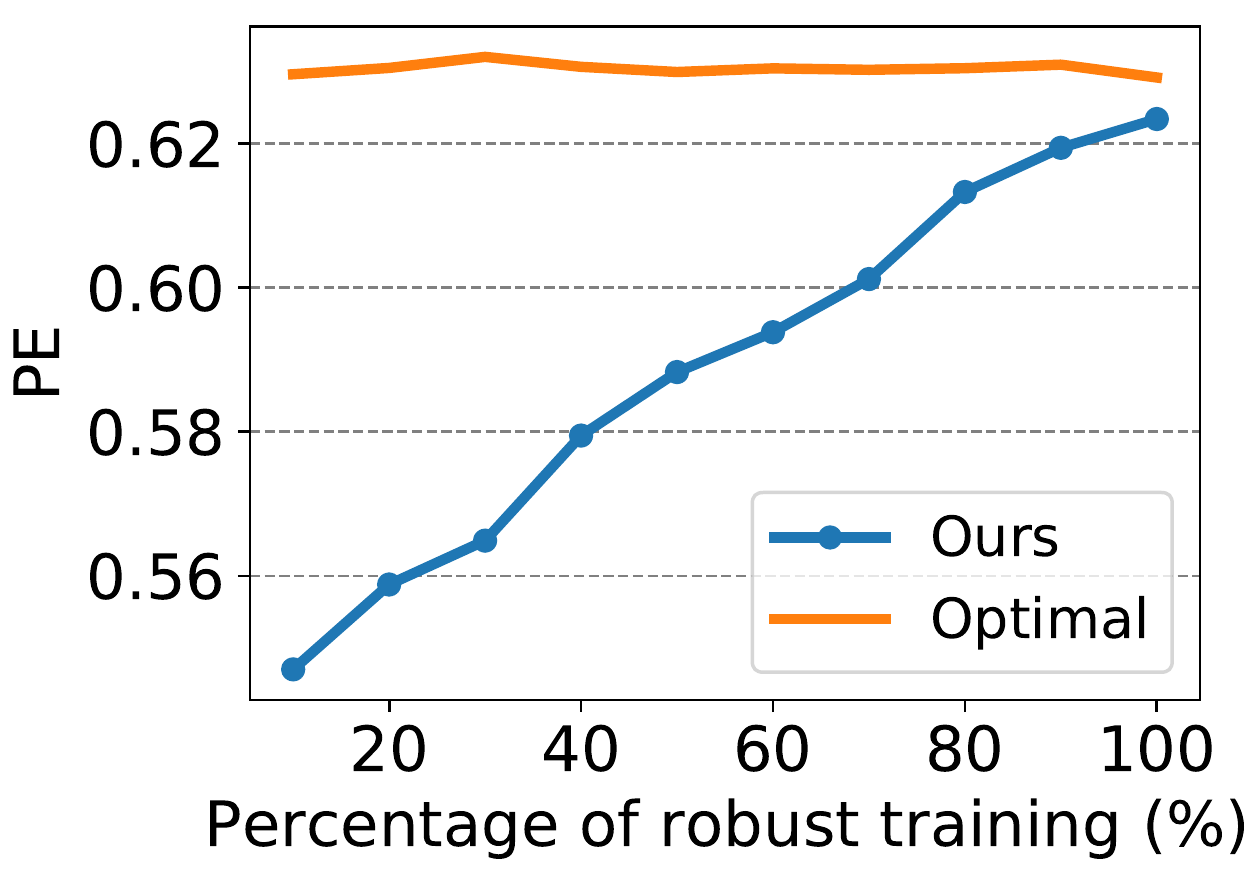}
    \captionof{figure}{Performance under robust defense}
    \label{fig:result_robust_range}
  \end{minipage}
\end{figure*}

\begin{figure*}[!t] 
\captionsetup{justification=centering}
  \begin{minipage}[t]{0.325\textwidth}
    \centering
    \includegraphics[width=2.3in]{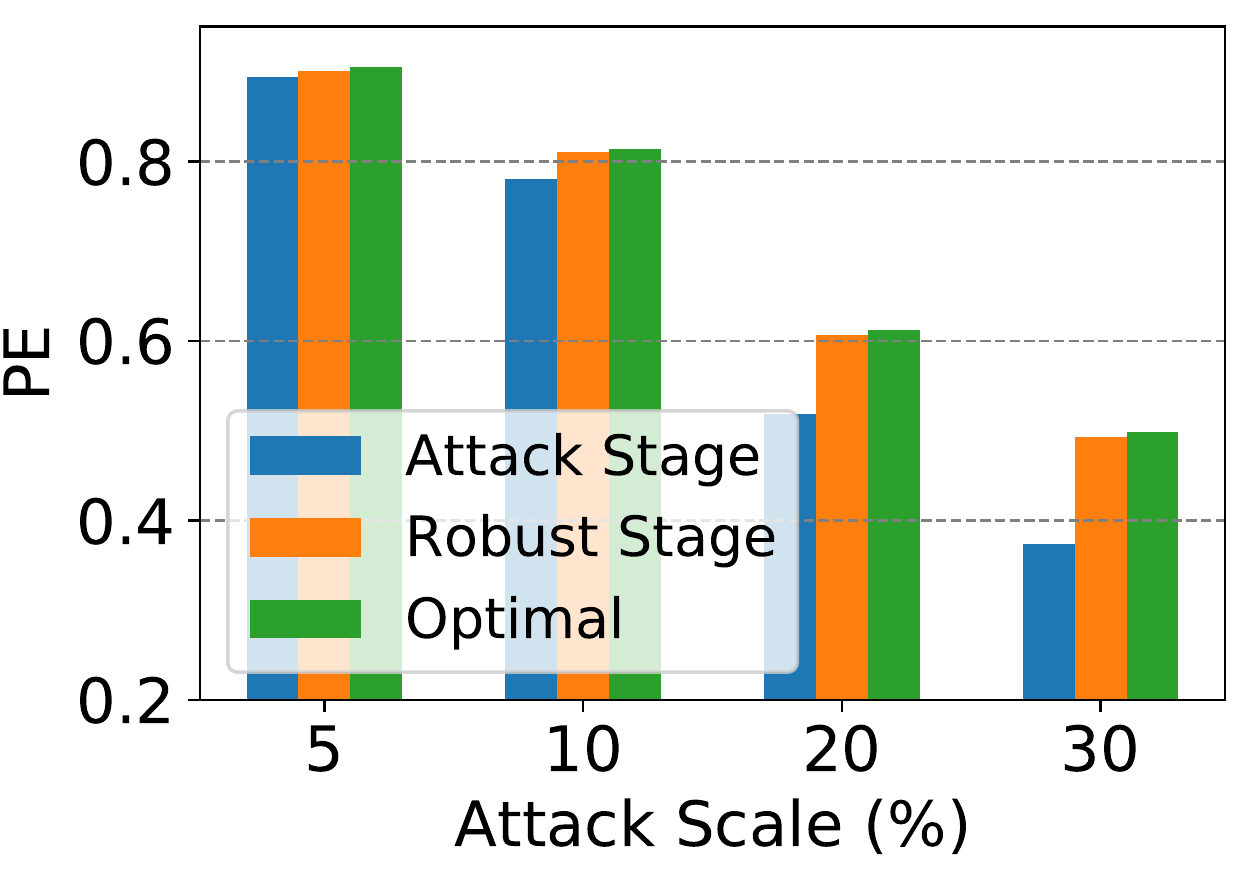}
    \captionof{figure}{Performance of \emph{RoNet} under different scales}
    \label{fig:result_noise_scale}
  \end{minipage}
  \hfill
  \begin{minipage}[t]{0.325\textwidth}
    \centering
    \includegraphics[width=2.3in]{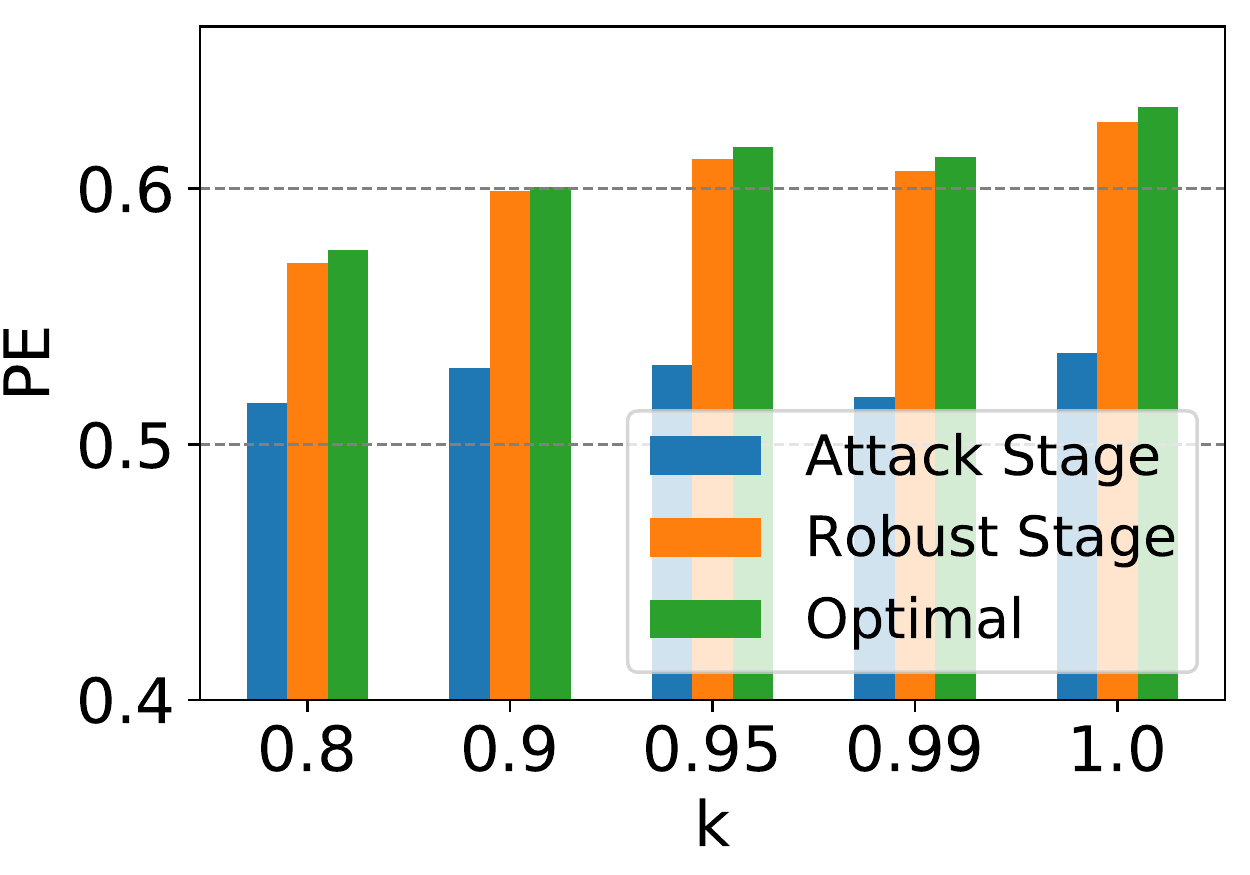}
    \captionof{figure}{Performance of \emph{RoNet} under different $\kappa$}
    \label{fig:result_percent}
  \end{minipage}
  \hfill
  \begin{minipage}[t]{0.325\textwidth}
    \centering
    \includegraphics[width=2.3in]{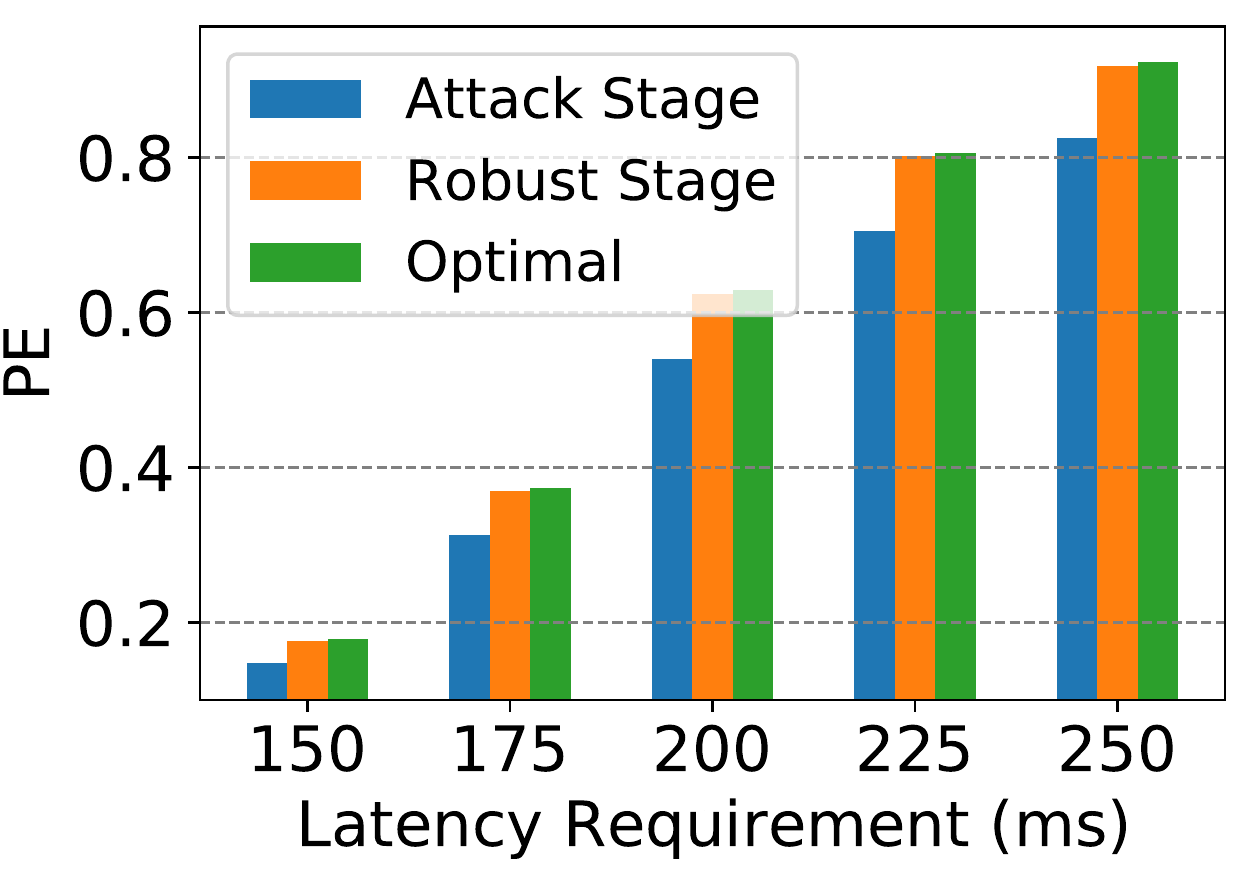}
    \captionof{figure}{Performance of \emph{RoNet} under different requirements}
    \label{fig:result_requirement}
  \end{minipage}
\end{figure*}

\vspace{-0.05in}
\section{Performance Evaluation}
\label{sec:evaluation}
In this section, we conduct extensive network simulations to evaluate the \emph{RoNet} framework in terms of efficacy, adaptability, and scalability.

\textbf{Network Simulator.}
We use the widely adopted Network Simulator 3 (NS-3)~\cite{ns3} to simulate the performance of mobile networks.
In particular, we develop an end-to-end network topology, including RAN, TN, Core, and Edge networks.
The RAN and Core are based on 4G LTE from the LENA project, where the channel model is \emph{LogDistancePropagationLossModel} and the total wireless bandwidth is 10MHz.
The \emph{avg\_distance} is randomized from 100 to 200 meters, and \emph{mcs\_max\_ul} and \emph{mcs\_max\_dl} are sampled from [4, 20] and [4, 28], respectively.
The TN is based on a p2p link with 1Gbps bandwidth and a 2 ms delay.
Specifically, we create a video analytics application for mobile users.
As simulations start, mobile users send uplink packets via the HTTP protocol with randomized \emph{ul\_avg\_size} between 10K and 20K Bytes.
We develop a new module to achieve queue-based edge computing simulations for processing the received packets.
The computation delay under 1.0 \emph{cpu\_ratio} is set to be 81 ms mean and 35 ms std, according to experimental measurements~\cite{liu2021onslicing}.
Without loss of generality, the downlink packet size is also randomly generated between 10K and 20K Bytes.
We design the simulation time to be 30 seconds to collect sufficient measurements of round-trip latency, which corresponds to average 21 seconds real-world time.
We empirically select the latency threshold as $H$=200 ms according to the capacity of the simulated network.
Note that \emph{RoNet} does not make assumptions for the network and can seamlessly adapt to other network settings and topologies.


\textbf{Parameters.}
We collect the dataset from the simulator with grid search on both 108 unique actions and 243 unique states, where the total number of state-action-latency transitions is 26244.
The data collection consumes more than 150 effective hours.
The dataset is used to estimate the PE under different states and actions during the simulation.
The architecture of the DNN model is [128]x[256]x[128] full-connected layer and ReLU activation function in PyTorch 1.10.
The PE in the evaluation is calculated by averaging the achieved PE under all states.
The default value of parameters are $\kappa$=0.99, $\epsilon$=0.2, and the number of training epochs is 50.

\textbf{Comparison.}
We implement the following methods for fair comparison with the \emph{RoNet} framework:
\begin{itemize}
    \item \emph{L-REG}: The L-REG uses the linear regression model to approximate the PE performance in Eq.~\ref{eq:PE}, samples actions and selects the one with the maximum predicted PE under individual states.
    \item \emph{R-REG}: The R-REG uses the Gaussian Process regression (GPR) model (implemented with \emph{scikit-learn} toolkit) to learn the PE performance, where the action selection is the same as that of L-REG.
    \item \emph{Optimal}: The Optimal is achieved by assuming that the attack is known, i.e., the attacked states can be observed. Then, the optimal action is obtained by extensive sampling the action space with the ground-truth dataset.
    \item \emph{RN}: The RN is the baseline of attacks, which randomly samples attacks from the attack space and selects the attack with the lowest predicted PE.
\end{itemize}

\textbf{Performance of Normal Training.} 
Fig.~\ref{fig:result_normal_range} shows the performance of different methods during the normal training stage.
As more training epochs are completed, the performance of \emph{RoNet} converges with the close approximation to the Optimal.
As compared to L-REG, R-REG can obtain higher PE, which is mainly attributed to the better approximation performance of Gaussian process regression (GPR). 
The \emph{RoNet} achieves nearly 12.5\% PE improvement than that of L-REG, which suggests the high efficacy of the deep neural network (DNN) based approximation.
This performance difference is expected to enlarge under higher dimensions of state and action space.

\textbf{Performance of Learn-to-Attack.} 
Fig.~\ref{fig:result_attack_range} depicts the performance of different attack methods during the learn-to-attack stage.
We observe that the PE can be dramatically decreased by attacking the state space, which indicates the vulnerability of the DNN-based configuration policy.
The mean and std of attack are \emph{avg\_distance} = 0.06/0.13, \emph{dl\_avg\_size} = 0.16/0.06, \emph{mcs\_max\_dl} = -0.1/0.15, \emph{mcs\_max\_ul} = -0.09/0.16, and \emph{ul\_avg\_size} = 0.01/0.17.
The high std indicates that attacks are contextual and individualized to different states, instead of static values.
As compared to RN, the learn-to-attack in \emph{RoNet} can achieve higher performance degradation (8.5\%), which justifies its effectiveness. 

\textbf{Performance of Robust Defense.} 
Fig.~\ref{fig:result_robust_range} shows the performance of \emph{RoNet} during the robust defense stage.
Here, we separate all the states into multiple subgroups, and continuously retrain the DNN to approximate the attacked PE performance on different subgroups.
As more state subgroups are applied for retraining, the PE performance of \emph{RoNet} recovers 13.9\% PE performance (from 0.547 to 0.623) and approaches the Optimal (0.629) gradually.
The performance of the Optimal is much lower than that in the normal training stage, because attacks are enforced on states.
We observe the attacked states are with lower MCS and higher uplink and downlink data sizes, where the performance cannot be fully recovered as more data to transmit and less MCS can be used.
As shown in Table~\ref{tb:performance}, when the robust defense stage completes, \emph{RoNet} achieves 0.623 PE which is very close to the Optimal (0.629).
As compared to L-REG, \emph{RoNet} can obtain more than 24.3\% improvement in the PE performance.
\begin{table}[!t]
\centering
    \begin{tabular}[b]{c|c|c}\hline
       \textbf{ }                       &  \textbf{Normal Performance} &  \textbf{Final Attacked Performance} \\ \hline
       \textbf{ Optimal}         & 0.985  & 0.629 \\ 
       \textbf{ Ours}           &  0.982 &  0.623\\ 
       \textbf{ G-REG}           &  0.958 &  0.543\\ 
       \textbf{ L-REG}           &  0.873 &  0.501\\ \hline
    \end{tabular}
    \captionof{table}{The performance of methods}
\label{tb:performance}
\end{table}

\textbf{Performance under Variabilities.}
Fig.~\ref{fig:result_noise_scale} shows the \emph{RoNet} performance under different attack scales.
In general, the higher attack scale allows larger attacks on the state space, and results in larger performance degradation.
When the attack scale is 0.05 and 0.1, the original PE performance (0.985) can be degraded to 0.90 (8.1\%) and 0.81 (17.8\%), respectively.
This disproportional performance degradation emphasizes the necessity of investigating the robustness of DNN-based configuration policy.
Fig.~\ref{fig:result_percent} shows the final attacked performance of \emph{RoNet} under different probabilistic action selection factor $\kappa$.
When $\kappa$ is 1, the action will always be selected if it has the maximum predicted PE among all sampled actions.
As we decrease $\kappa$, the set of action candidates expands, which generally improves the robustness because the attacker has to target more actions.
Consequently, the PE performance is decreased because of the random selection on the set of suboptimal action candidates.
Fig.~\ref{fig:result_requirement} shows the impact of latency threshold $H$ on the PE performance in \emph{RoNet}.
With the more stringent latency threshold, the PE decreases significantly under the fixed transmission and computation resources in the simulated network.

In the default parameters, the mean and std of actions are $bandwidth\_ul$ = 0.56/0.1, $bandwidth\_dl$ = 0.66/0.18, and $cpu\_ratio$ = 0.98/0.03.
To counter the unknown attacks on the state space, the mean final actions are $bandwidth\_ul$ = 0.7/0.17, $bandwidth\_dl$ = 0.79/0.2, and $cpu\_ratio$ = 0.99/0.01.
It is worth noting that \emph{RoNet} can always approach the Optimal under different variabilities, which justifies its strong adaptability and generalization.

\vspace{-0.05in}
\section{Related Work}

\textbf{Network configuration} has been extensively studied in mobile networks to optimize a variety of networking objectives.
Liu \emph{et. al.}~\cite{liu2018joint} formulated an approximated math model to describe the problem of radio and computation resource management, and designed an iterative algorithm to search for the optimal solution for minimizing the latency of mobile users.
There are other works~\cite{ran2018deepdecision, d2020sl, salvat2018overbooking, d2022orchestran} build their analytic model based on experimental measurements for different applications, e.g., augmented reality~\cite{ran2018deepdecision} and network slicing~\cite{salvat2018overbooking}.
Recent works~\cite{zhang2020onrl, shi2021adapting, liu2021onslicing} focus on machine learning approaches to learn and automate network configuration.
Shi \emph{et. al.}~\cite{shi2021adapting} exploited deep learning to learn and configure wireless mesh networks and used domain adaptation techniques to bridge the discrepancy between simulators and real-world networks.  
However, these ML-based solutions fail to effectively counter real-world variabilities and potential attacks in mobile networks. 

\textbf{AI robustness} has received increasing research attention, especially for DNN-based policies in real-world deployments.
In the pioneer AI robustness work~\cite{madry2017towards}, the vulnerability of DNN under adversarial attacks is revealed with extensive experiments, where the authors proposed a first-order adversary method to achieve robust training.
Multiple derivative works aim to improve the robustness of DNN and deep reinforcement learning (DRL) by using diversified techniques, e.g., data augmentation~\cite{rebuffi2021fixing}, sparsity architecture~\cite{guo2018sparse}, and local linearization~\cite{qin2019adversarial}.
However, these works focused on individual DNN model with the strong assumption of known gradients during the DNN training.
In this work, we aim to derive a robust configuration policy, including robust DNN-based approximation and action selections.

\vspace{-0.08in}
\section{Conclusion}
In this work, we presented a new \emph{RoNet} framework with three stages in mobile networks to improve the robustness of configuration policies.
First, we designed a neural-assisted network configuration policy that maximizes the performance efficiency of mobile users.
Second, we designed a learn-to-attack method to automatically attack the state observation and degrade the performance of the policy.
Third, we designed a robust defense method to recover the under-attack performance of the policy.
The performance evaluation shows that \emph{RoNet} outperforms existing solutions in terms of robustness, adaptability, and scalability. 

\section*{Acknowledgement}
This work is partially supported by the US National Science Foundation under Grant No. 2212050.

\vspace{-0.05in}
\bibliographystyle{IEEEtran}
\bibliography{ref/reference, ref/qiang}

\end{document}